\documentstyle[aps,pre,preprint]{revtex}
\tightenlines
\begin{document}
\draft
\title{The Long--Term Future of Extragalactic Astronomy}
\author{Abraham Loeb}
\address{Astronomy Department, Harvard University, 60 Garden Street,
  Cambridge, MA 02138; aloeb@cfa.harvard.edu}
\date{\today}
\maketitle

\begin{abstract} 
If the current energy density of the universe is indeed dominated by a
cosmological constant, then high--redshift sources will remain visible to
us only until they reach some finite age in their rest--frame. The
radiation emitted beyond that age will never reach us due to the
acceleration of the cosmic expansion rate, and so we will never know what
these sources look like as they become older. As a source image freezes on
a particular time frame along its evolution, its luminosity distance and
redshift continue to increase exponentially with observation time.  The
higher the current redshift of a source is, the younger it will appear as
it fades out of sight.  For the popular set of cosmological parameters, I
show that a source at a redshift $z_0\sim 5$--10 will only be visible up to
an age of $\sim 4$--$6$ billion years. Arguments relating the properties of
high--redshift sources to present--day counterparts will remain indirect
even if we continue to monitor these sources for an infinite amount of
time. These sources will not be visible to us when they reach the current
age of the universe. 
\end{abstract}
\input epsf.sty

\section{Introduction}
Recent observations of the microwave background and Type Ia supernovae
indicate that the universe is flat and its expansion is currently dominated
by a cosmological constant
\cite{Garnavich1998,Perlmutter1999,deBernardis2000,Hanany2000,Netterfield2001,Lee2001,Wang2001}.
The cosmological scale factor may be just entering an exponential expansion
phase (similar to inflation) during which all comoving observers will
eventually lose causal contact with each other.

The qualitative implications of the future exponential expansion were
discussed in the literature \cite{Starobinsky2000,Krauss2000}. It was
recognized that an asympototically de Sitter universe of this type
possesses a moving light cone out of which all distant sources will
eventually exit \cite{Starkman1999,Gudmundsson2001}.  In similarity to the
case of a black hole, one can define an event horizon \cite{Rindler1956}
out to which events can in principle be seen by us (for definitions of
other surfaces see \cite{Rindler1956,Other_horizons,Gudmundsson2001}).  In
an asympototically de Sitter universe, the event horizon asymptotes to a
fixed proper distance from us\cite{Rindler1956,Gudmundsson2001} above which
cosmological events will never be visible to us. Due to the exponential
expansion of the cosmological scale factor, all sources which follow the
Hubble expansion will eventually exit (in their own reference frame) out of
our event horizon.  The further a source is away from us, the earlier it
exits.

But since our universe was matter-dominated earlier in its history, the
number of cosmological sources visible to us has been increasing steadily
with cosmic time until recently. It is therefore interesting to examine
quantitatively what will happen in the future to the images of all the
currently visible sources as a function of their current redshifts.  In
this paper I show quantitatively that as a result of the acceleration in
the cosmic expansion, all high-redshift sources will fade out of our sight
at a finite age (similarly to a source which is infalling through the
horizon of a black hole).  This implies that we will never be able to see
their image as they get older. In \S 2, I calculate the {\it maximum
visible age} of a cosmological source as a function of its currently
measured redshift. For concreteness, I adopt the present--day density
parameter values of $\Omega_M=0.3$ for matter and $\Omega_\Lambda=0.7$ for
the cosmological constant.

\section{Maximum Visible Age as a Function of Current Source Redshift}
The line-element for a flat universe is given by $ds^2=c^2 dt^2 -
a^2(t)(dr^2+ r^2 d\Omega)$, where $a(t)$ is the scale factor. Photon
trajectories satisfy $ds=0$, and so the comoving distance of a source that
emits radiation at a cosmic time $t_{\rm em}$ and is observed at the
current age of the universe $t_0$ is given by,
\begin{equation}
r =\int_{t_{\rm em}}^{t_0} {c~dt\over a(t)}
\label{eq:rt}
\end{equation}
If the source continues to emit at a later time $t_{\rm em}^\prime$, then
this radiation will be observed by us at a future time
$t_{0}^\prime$. Since the source maintains its comoving coordinate, 
\begin{equation}
r= \int_{t_{\rm em}}^{t_0} {c~dt \over a(t)}=\int_{t_{\rm
em}^\prime}^{t_0^\prime} {c~dt\over a(t)},
\end{equation}
or equivalently
\begin{equation}
\int_{t_0}^{t_0^\prime} {dt\over a(t)}=\int_{t_{\rm em}}^{t_{\rm
em}^\prime} {dt\over a(t)}.
\label{eq:basic}
\end{equation}
In terms of the conformal time, $\eta (t)\equiv \int_0^t dt^\prime
/a(t^\prime)$, equation~(\ref{eq:basic}) is equivalent to the condition
$(\eta (t_0^\prime)-\eta (t_0))=(\eta (t_{\rm em}^\prime)-\eta (t_{\rm
em}))$.  The question of whether this equality can be satisfied for an
arbitrary value of the source age, $t_{\rm em}^\prime$, depends on the
future evolution of the scale factor $a(t)$. It is easy to see that as long
as $0<d \ln a/ d t< 1$, this equality can be satisfied for an arbitrary
value of $t_{\rm em}^\prime$. This is the case, for example, in a
matter-dominated universe where $a\propto t^{2/3}$. However, in a de Sitter
universe the scale factor grows exponentially and so the integrand on the
left-hand-side of equation~(\ref{eq:basic}) saturates at a finite value
even as $t_0^\prime \rightarrow \infty$.  This implies that there is a
maximum intrinsic age, $t_{\rm em}^\prime$, over which the source is
visible to us.  Emission after the source reaches this age will never be
observable by us\footnote{The observed redshift of the source diverges
exponentially as $t_0^\prime \rightarrow \infty$ and so does the luminosity
distance (see Fig. 3). Hence, the flux received from the source declines
exponentially with increasing observing time $t_0^\prime$. As the image of
the source fades away it stays frozen at a fixed time along its evolution.
This situation is qualitatively analogous to the observed properties of a
source falling through the event horizon of a Schwarzschild black hole
\cite{Shapiro1983}.} (unless the vacuum energy density which makes up the
cosmological constant decays). The maximum visible age obviously depends on
$t_{\rm em}$ or the currently measured source redshift, $z_0$, which is
given by the relation $a(t_{\rm em})=(1+z_0)^{-1}$.

The evolution of the scale factor is determined by the Friedmann equation,
\begin{equation}
{1\over a}{da \over dt}= H_0 \left({\Omega_M \over a^3}+ \right.
\left. \Omega_\Lambda\right)^{1/2},
\label{eq:dadt}
\end{equation}
where $\Omega_M+\Omega_\Lambda=1$. Equations~(\ref{eq:rt}) and
(\ref{eq:dadt}) admit analytic solutions 
\cite{Edwards1972,Edwards1973,Dabrowski1986,Dabrowski1987,Weinberg1972,Weinberg1989,Eisenstein1997}
for $r(t_{\rm em},t_0)$ in terms of an incomplete elliptic integral,
and for the scale factor in the form
\begin{equation}
a(t)=\left({\Omega_M \over 1- \Omega_M}\right)^{1/3} \left( {\rm
sinh}\left({3\over 2}\sqrt{1-\Omega_M} H_0 t\right)\right)^{2/3}.
\end{equation}
Pen \cite{Pen1999} provides a simple fitting formula for
$\eta(a)=(r(0)-r(a))$.  The luminosity and angular diameter distances at
any future time $t_0^\prime$ are given by
$d_L=\{a^2(t_0^\prime)/a(t_1^\prime)\}r$ and $d_A=a(t_1^\prime)r$,
respectively \cite{Weinberg1972}. The source redshift evloves as
$z=\{a(t_0^\prime)/a(t_1^\prime)\}-1$.

Figure 1 shows the emission time, $t_{\rm em}^\prime$, as a function of the
future observing time, $t_0^\prime$. All time scales are normalized by the
inverse of the current Hubble expansion rate, $H_0=(\dot{a}/a)\vert_{t_0}$.
Clearly, as the current source redshift increases, its maximum visible age
in the future (i.e. the asymptotic value of $t_{\rm em}^\prime$ for
$t_0^\prime \rightarrow \infty$) decreases. Typically, the maximum emission
time $t_{\rm em}^\prime$ is much longer than the current emission time
$t_{\rm em}$, and so only sources that are steady over many Hubble times at
their current redshift are suitable for this discussion.

The microwave background anisotropies, for example, do not possess the
above property since they were generated over a narrow temporal interval
around the time of recombination, $t_{\rm rec}$ (corresponding to $z_0\sim
1000$). Hence, the comoving distance of their last scattering surface will
increase with the advance of cosmic time, $r_{\rm
rec}(t_0^\prime)=\int_{t_{\rm rec}}^{t_0^\prime} dt/a(t)$, and we will be
seeing spatial regions that were more distant from us at $t_{\rm rec}$.
Eventually, $r_{\rm rec}$ will approach a constant value $\sim 4.4
cH_0^{-1}$ at $t_0^\prime \gtrsim 4H_0^{-1}$ and the background anisotropy
pattern on the sky will freeze. Since the comoving scale associated with
the first acoustic peak of the anisotropies is $\sim 100 h_{0.7}^{-1}$ Mpc
and the asymptotic value of $r_{\rm rec}$ is different from its preset
value by $1.14 c H_0^{-1}= 4.9 h_{0.7}^{-1}$ Gpc, we will be able to sample
only $\sim 50$ independent realizations of the density fluctuation mode
corresponding to the first peak. This implies that the cosmic variance of
the first acoustic peak would be at best reduced by a factor of $\sim
\sqrt{50}=7.1$ relative to its value today. The statistics improve, of
course, for modes with shorter wavelengths.

The upper panel of Figure 2 shows the maximum visible age of a source
(starting from the Big Bang) as a function of its currently measured
redshift. The lower panel gives the corresponding redshift below which it
will not be possible to identify a counterpart to the source in a current
deep image of the universe, even if we continue to monitor this source
indefinitely.

As the source image freezes on a particular time frame along its evolution
(Fig. 1), its flux continues to decline and its redshift increases. Figure
3 shows the evolution of the luminosity and angular diameter distances
(relative to their values today) as well as the source redshift as
functions of observation time, for a source with a present--day redshift 
$z_0=5$. Although $d_L\propto \exp (2\sqrt{1-\Omega_M}H_0t)$ and $z\propto
\exp (\sqrt{1-\Omega_M}H_0t)$ diverge exponentially at $t_0^\prime\gg t_0$,
the angular diameter distance $d_A$ approaches a constant value. A source
with a constant intrinsic size at $z_0=5$ will occupy in the distant future
a fixed angular size on the sky, which is $\sim 3.3$ times larger than its
angular size today.

\section{Discussion}

The quantitative results of this work are summarized in Figures 1--3.  In
an asymptotically de Sitter universe, we can see sources up to the time
(given in Fig. 1) when they crossed our event horizon.  Figure 2 implies
that a source at a redshift $z_0=5$ will only be visible to us up to an age
of $\sim 6.4 h_{0.7}^{-1}$ Gyr.  Thus, we will never be able to observe the
evolution of this source and identify its counterpart in a map that we have
taken today of the universe at a redshift $z_0< 0.8$, even if we continue
to monitor this source indefinitely.  This is because the age of the
currently observed universe at $z_0\lesssim 0.8$ exceeds $6.6 h_{0.7}^{-1}$
Gyr.  In other words, arguments relating the properties of high-redshift
sources to counterparts in the present--day universe will forever remain
indirect. Similarly, any light signal that we send out today will not be
able to reach all sources with current redshifts $z_0\gtrsim 1.8$ (see
Fig. 2).

The {\it visible age} limit becomes stricter for flux--limited observations
where the maximum value of $t_0^\prime$ is constrained by the requirement
that the luminosity distance will not exceed some value (see Fig. 3). While
the flux limit may depend on technological advances in instrumentation, the
visible age limit derived in this paper for $t_0^\prime\rightarrow \infty$
is absolute.

The results illustrated in Figures 1--3 might change only if the vacuum
energy density, $\rho_{\rm V}$, which makes up the cosmological constant
would decrease significantly over the next few Hubble times or $\sim
5\times 10^{10}~{\rm years}$ \cite{Starobinsky2000,Barrow2000}. The
exponential expansion phase will not occur if the vacuum energy density
would eventually vanish.  Although this behaviour is possible in the case
of a rolling scalar field or ``Quintessence''
\cite{Ratra1988,Caldwell1998}, it requires that the equation of state of
the corresponding ``dark energy'' would deviate significantly from the
$p_{\rm V}=-\rho_{\rm V} c^2$ relation that characterizes the pressure
$p_{\rm V}$ of a true cosmological constant. A past deviation as small as
$\lesssim 10\%$ from this relation is measurable by forthcoming projects,
such as the proposed Supernova/Acceleration Probe (SNAP)
mission\footnote{http://snap.lbl.gov/} which intends to monitor $\lesssim
2000$ Type Ia supernovae across the sky per year and determine their
luminosity distances up to a redshift $z_0\sim 1.5$ with high precision.

As long as $\rho_{\rm V}$ will remain nearly constant, the prospects for
extragalactic astronomy in the long--term future appear grim\footnote{As
far as individual objects (such as planets, stars, or galaxies) are
concerned, their evolution into the much longer term future of the universe
has been discussed in detail in the literature (see Ref. \cite{Adams1997}
and references therein).}.  In contrast to a matter-dominated universe
\cite{Rothman1987}, the statistics of visible sources in a
$\Lambda$-dominated universe are getting worse with the advance of cosmic
time.  Within $\lesssim 10^{11}$ years, we will be able to see only those
galaxies that are gravitationally bound to the Local Group of galaxies,
including the Virgo cluster and possibly some parts of the local
supercluster (where the global overdensity in a sphere around Virgo is
larger than a few). All other sources of light will fade away beyond
detection and their fading image will be frozen at a fixed age.

\bigskip
\bigskip
\paragraph*{Acknowledgments.}

The author thanks Rennan Barkana, George Rybicki, and Matias Zaldarriaga
for useful discussions.  This work was supported in part by NASA grants NAG
5-7039, 5-7768, and by NSF grants AST-9900877, AST-0071019.

\newpage

\begin{figure}
\epsfxsize=6.0in\epsfbox{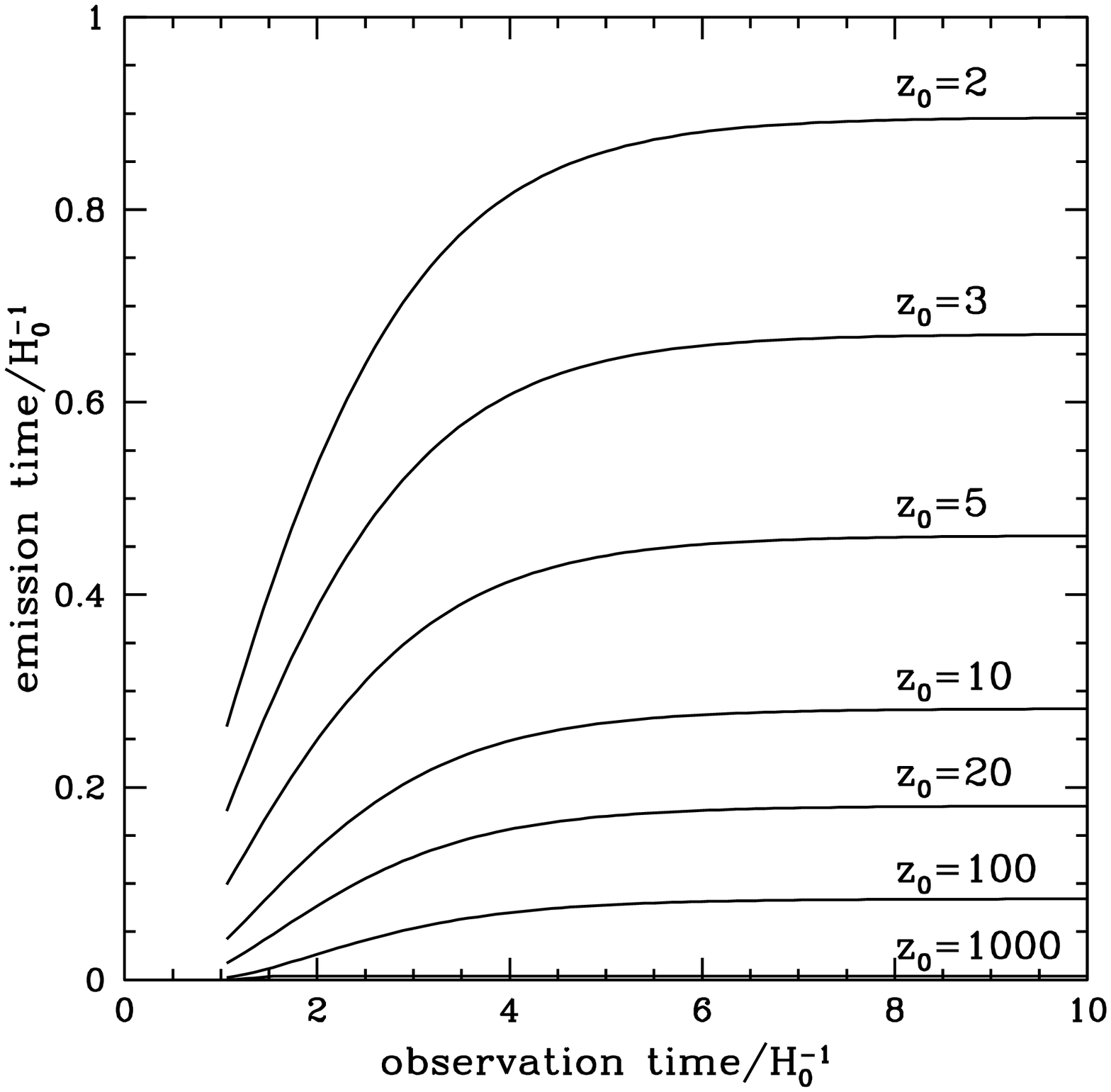} 
\bigskip
\bigskip
\bigskip
\caption{Emission time as a function of future
observation time for an $\Omega_M=0.3$, $\Omega_\Lambda=0.7$ universe. Time
is measured in units of $H_0^{-1}=14 h_{0.7}^{-1}~{\rm Gyr}$, where
$h_{0.7}\equiv (H_0/70~{\rm km~s^{-1}~Mpc^{-1}})$. The current time is
$t_0=0.96 H_0^{-1}$. For any currently measured redshift $z_0$ of a source,
there is a maximum intrinsic age up to which we can see that source even if
we continue to monitor it indefinitely.  }
\label{Fig1}
\end{figure}

\newpage

\begin{figure}
\epsfxsize=6.0in\epsfbox{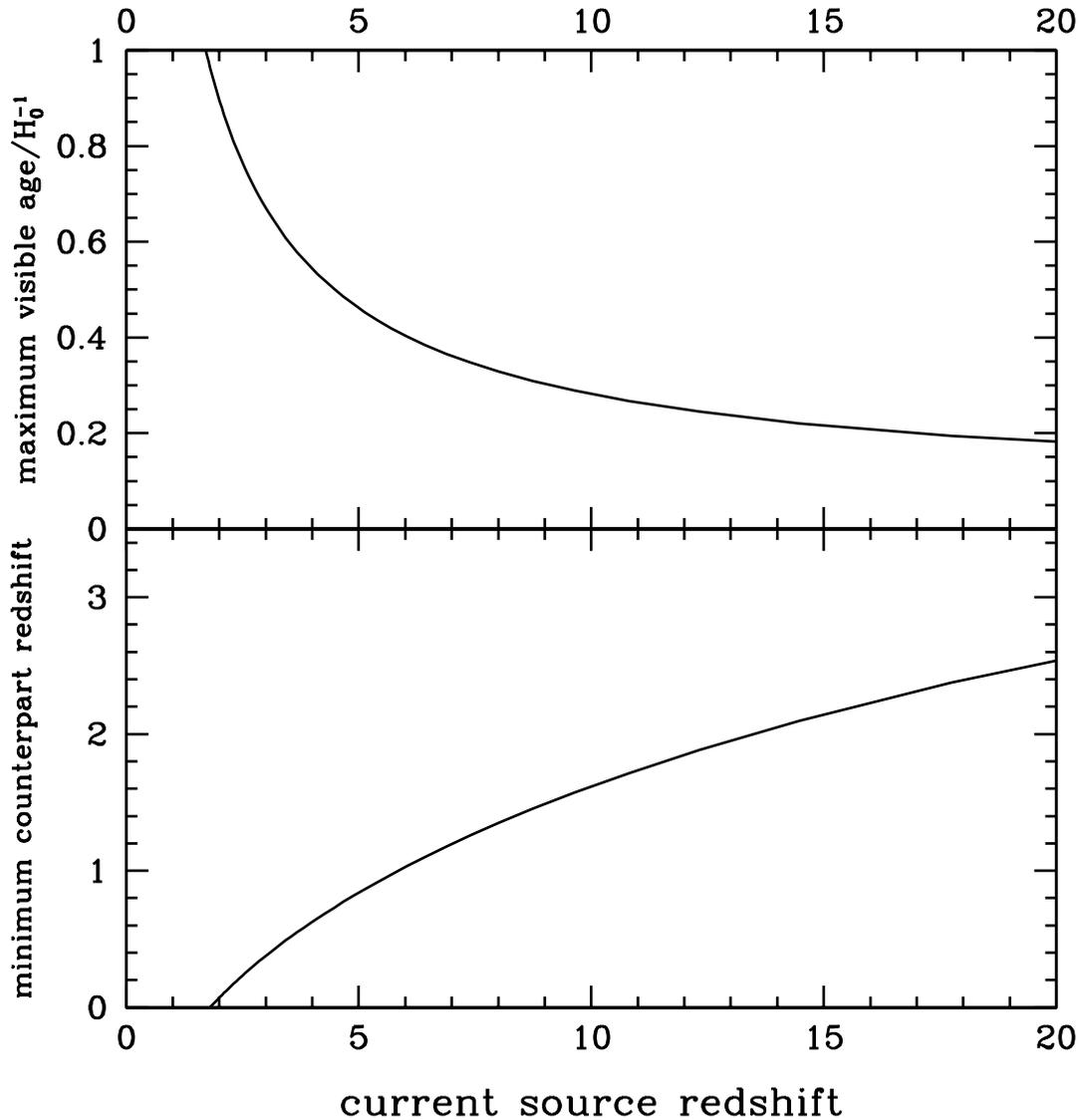} 
\bigskip
\bigskip
\bigskip
\caption{The upper panel shows the maximum {\it
visible age} of a source (in units of $H_0^{-1}=14 h_{0.7}^{-1}~{\rm Gyr}$)
as a function of its currently measured redshift, $z_0$. The lower panel
shows the redshift at which the age of the universe equals this maximum
{\it visible age} of the source. This is the minimum redshift for which it
will be possible, in principle, to identify a counterpart to the source in
a current deep image of the universe. The counterparts of all sources at
$z_0< 1.8$ can be traced to the present time.}
\label{Fig2}
\end{figure}

\newpage

\begin{figure}
\epsfxsize=5.0in\epsfbox{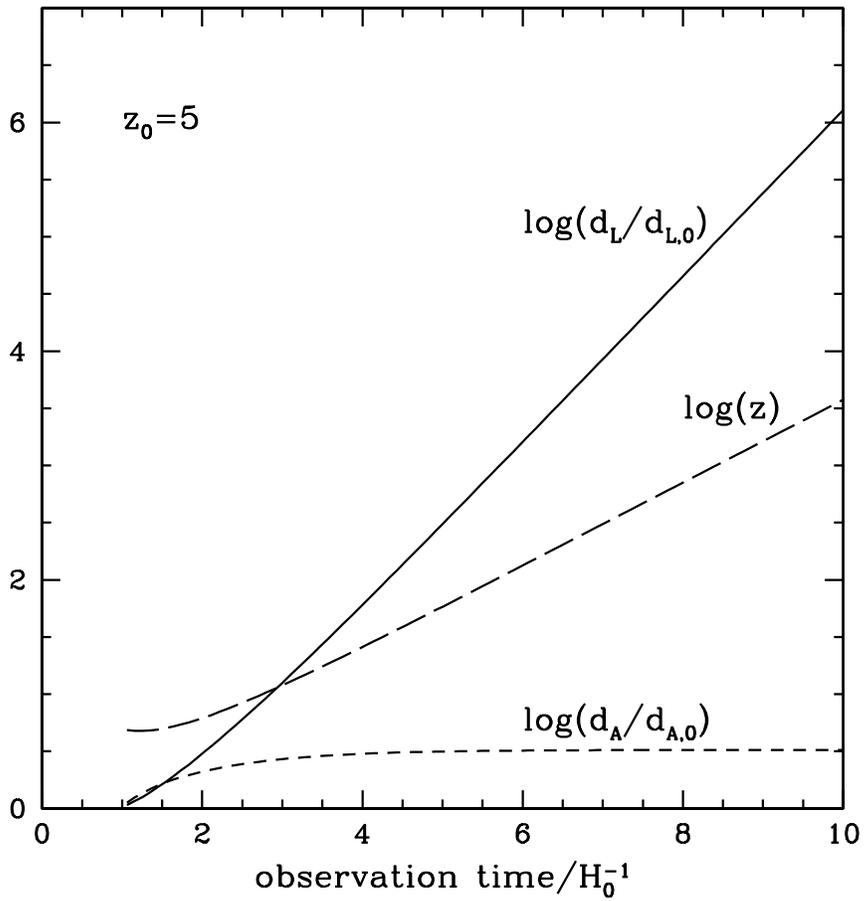} 
\bigskip
\bigskip
\bigskip
\caption{Future evolution of the luminosity and angular diameter distances
[$d_L(t_0^\prime)$, $d_A(t_0^\prime$)] relative to their values today
[$d_{L,0}\equiv d_L(t_0)$, $d_{A,0}\equiv d_A(t_0)$] and the observed
redshift $z(t_0^\prime)$ for a source with a present--day redshift
$z_0=5$.}  \label{Fig3} 
\end{figure} 
\end{document}